

\font\twelverm=cmr10 scaled 1200    \font\twelvei=cmmi10 scaled 1200
\font\twelvesy=cmsy10 scaled 1200   \font\twelveex=cmex10 scaled 1200
\font\twelvebf=cmbx10 scaled 1200   \font\twelvesl=cmsl10 scaled 1200
\font\twelvett=cmtt10 scaled 1200   \font\twelveit=cmti10 scaled 1200

\skewchar\twelvei='177   \skewchar\twelvesy='60
\def\twelvepoint{\normalbaselineskip=12.4pt
  \abovedisplayskip 12.4pt plus 3pt minus 9pt
  \belowdisplayskip 12.4pt plus 3pt minus 9pt
  \abovedisplayshortskip 0pt plus 3pt
  \belowdisplayshortskip 7.2pt plus 3pt minus 4pt
  \smallskipamount=3.6pt plus1.2pt minus1.2pt
  \medskipamount=7.2pt plus2.4pt minus2.4pt
  \bigskipamount=14.4pt plus4.8pt minus4.8pt
  \def\rm{\fam0\twelverm}          \def\it{\fam\itfam\twelveit}%
  \def\sl{\fam\slfam\twelvesl}     \def\bf{\fam\bffam\twelvebf}%
  \def\mit{\fam 1}                 \def\cal{\fam 2}%
  \def\tt{\twelvett}
  \textfont0=\twelverm   \scriptfont0=\tenrm   \scriptscriptfont0=\sevenrm
  \textfont1=\twelvei    \scriptfont1=\teni    \scriptscriptfont1=\seveni
  \textfont2=\twelvesy   \scriptfont2=\tensy   \scriptscriptfont2=\sevensy
  \textfont3=\twelveex   \scriptfont3=\twelveex  \scriptscriptfont3=\twelveex
  \textfont\itfam=\twelveit
  \textfont\slfam=\twelvesl
  \textfont\bffam=\twelvebf \scriptfont\bffam=\tenbf
  \scriptscriptfont\bffam=\sevenbf
  \normalbaselines\rm}

\def\beginlinemode{\endmode
  \begingroup\parskip=0pt \obeylines\def\\{\par}\def\endmode{\par\endgroup}}
\def\beginparmode{\endmode
  \begingroup \def\endmode{\par\endgroup}}
\let\endmode=\par
{\obeylines\gdef\
{}}
\def\singlespace{\baselineskip=\normalbaselineskip}

\def\oneandahalfspace{\baselineskip=\normalbaselineskip
  \multiply\baselineskip by 3 \divide\baselineskip by 2}
\def\doublespace{\baselineskip=\normalbaselineskip \multiply\baselineskip by 2}

\newcount\firstpageno
\firstpageno=2
\footline={\ifnum\pageno<\firstpageno{\hfil}\else{\hfil\twelverm\folio\hfil}\fi}
\let\rawfootnote=\footnote		
\def\footnote#1#2{{\rm\singlespace\parindent=0pt\rawfootnote{#1}{#2}}}
\def\raggedcenter{\leftskip=4em plus 12em \rightskip=\leftskip
  \parindent=0pt \parfillskip=0pt \spaceskip=.3333em \xspaceskip=.5em
  \pretolerance=9999 \tolerance=9999
  \hyphenpenalty=9999 \exhyphenpenalty=9999 }
\hsize=6.5truein
\vsize=8.9truein
\parskip=\medskipamount
\twelvepoint		
\doublespace		
\overfullrule=0pt	
\def\preprintno#1{
 \rightline{\rm #1}}	

\def\title			
  {\null\vskip 3pt plus 0.2fill
   \beginlinemode \doublespace \raggedcenter \bf}

\def\author			
  {\vskip 4pt plus 0.2fill \beginlinemode
   \singlespace \raggedcenter}

\def\affil			
  {\vskip 3pt plus 0.1fill \beginlinemode
   \oneandahalfspace \raggedcenter \sl}

\def\abstract			
  {\vskip 3pt plus 0.3fill \beginparmode
   \doublespace \narrower }

\def\endtitlepage		
  {\endpage			
   \body}

\def\body			
  {\beginparmode}		

\def\head#1{			
  \filbreak\vskip 0.5truein	
  {\immediate\write16{#1}
   \line{\bf{#1}\hfil}\par}
   \nobreak\vskip 0.25truein\nobreak}

\def\refto#1{$^{#1}$}		

\def\references			
  {\head{References}		
   \beginparmode
   \frenchspacing \parindent=0pt \leftskip=0truecm
   \parskip=8pt plus 3pt \everypar{\hangindent=\parindent}}

\gdef\refis#1{\indent\hbox to 0pt{\hss#1.~}}	

\gdef\journal#1, #2, #3, 1#4#5#6{		
    {\sl #1~}{\bf #2}, #3, (1#4#5#6)}		

\def\refstylenp{		
  \gdef\refto##1{ [##1]}				
  \gdef\refis##1{\indent\hbox to 0pt{\hss##1)~}}	
  \gdef\journal##1, ##2, ##3, ##4 {			
     {\sl ##1~}{\bf ##2~}(##3) ##4 }}
\def\np{\journal Nucl. Phys., }
\def\endreferences{\body}
\def\endpage			
  {\vfill\eject}
\def\endpaper                   
  {\endmode\vfill\supereject}
\def\endit
  {\endpaper\end}
\def\ref#1{Ref.\thinspace#1}			

\def\Ref#1{Ref.\thinspace#1}			

\def\eq#1{Eq.\thinspace(#1)}			

\def\(#1){Eq.\thinspace(\call{#1})}
\def\call#1{{#1}}

\def\frac#1#2{{#1 \over #2}}
\def\cft{conformal field theory}
\def\cfts{conformal field theories}
\def\rg{renormalization group}
\def\ope{operator product expansion}
\def\const{{\rm const.}}
\def\a{\alpha}
\def\b{\beta}
\def\th{\theta}
\def\d{\delta}
\def\ev#1{\langle #1\rangle}
\def\D{\Delta}
\def\hb{\bar h}
\def\db{\bar\delta}
\def\e{\epsilon}
\def\t{\tau}
\def\zb{\bar z}
\def\Tb{\overline T}
\def\ffrac#1#2{\textstyle{#1\over#2}\displaystyle}
\def\rhs{right hand side}
\def\om{\omega}
\def\k{\kappa}
\def\bra#1{\langle #1\vert}
\def\ket#1{\vert #1\rangle}
\def\Wb{\overline W}

\refstylenp

\catcode`@=11
\newcount\r@fcount \r@fcount=0
\newcount\r@fcurr
\immediate\newwrite\reffile
\newif\ifr@ffile\r@ffilefalse
\def\w@rnwrite#1{\ifr@ffile\immediate\write\reffile{#1}\fi\message{#1}}

\def\writer@f#1>>{}
\def\referencefile{
  \r@ffiletrue\immediate\openout\reffile=\jobname.ref%
  \def\writer@f##1>>{\ifr@ffile\immediate\write\reffile%
    {\noexpand\refis{##1} = \csname r@fnum##1\endcsname = %
     \expandafter\expandafter\expandafter\strip@t\expandafter%
     \meaning\csname r@ftext\csname r@fnum##1\endcsname\endcsname}\fi}%
  \def\strip@t##1>>{}}

\def\citeall#1{\xdef#1##1{#1{\noexpand\cite{##1}}}}
\def\cite#1{\each@rg\citer@nge{#1}}	

\def\each@rg#1#2{{\let\thecsname=#1\expandafter\first@rg#2,\end,}}
\def\first@rg#1,{\thecsname{#1}\apply@rg}	
\def\apply@rg#1,{\ifx\end#1\let\next=\relax
\else,\thecsname{#1}\let\next=\apply@rg\fi\next}

\def\citer@nge#1{\citedor@nge#1-\end-}	
\def\citer@ngeat#1\end-{#1}
\def\citedor@nge#1-#2-{\ifx\end#2\r@featspace#1 
  \else\citel@@p{#1}{#2}\citer@ngeat\fi}	
\def\citel@@p#1#2{\ifnum#1>#2{\errmessage{Reference range #1-#2\space is bad.}%
    \errhelp{If you cite a series of references by the notation M-N, then M and
    N must be integers, and N must be greater than or equal to M.}}\else%
 {\count0=#1\count1=#2\advance\count1
by1\relax\expandafter\r@fcite\the\count0,%
  \loop\advance\count0 by1\relax
    \ifnum\count0<\count1,\expandafter\r@fcite\the\count0,%
  \repeat}\fi}

\def\r@featspace#1#2 {\r@fcite#1#2,}	
\def\r@fcite#1,{\ifuncit@d{#1}
    \newr@f{#1}%
    \expandafter\gdef\csname r@ftext\number\r@fcount\endcsname%
                     {\message{Reference #1 to be supplied.}%
                      \writer@f#1>>#1 to be supplied.\par}%
 \fi%
 \csname r@fnum#1\endcsname}
\def\ifuncit@d#1{\expandafter\ifx\csname r@fnum#1\endcsname\relax}%
\def\newr@f#1{\global\advance\r@fcount by1%
    \expandafter\xdef\csname r@fnum#1\endcsname{\number\r@fcount}}

\let\r@fis=\refis			
\def\refis#1#2#3\par{\ifuncit@d{#1}
   \newr@f{#1}%
   \w@rnwrite{Reference #1=\number\r@fcount\space is not cited up to now.}\fi%
  \expandafter\gdef\csname r@ftext\csname r@fnum#1\endcsname\endcsname%
  {\writer@f#1>>#2#3\par}}

\def\ignoreuncited{
   \def\refis##1##2##3\par{\ifuncit@d{##1}%
     \else\expandafter\gdef\csname r@ftext\csname
r@fnum##1\endcsname\endcsname%
     {\writer@f##1>>##2##3\par}\fi}}

\def\r@ferr{\endreferences\errmessage{I was expecting to see
\noexpand\endreferences before now;  I have inserted it here.}}
\let\r@ferences=\references
\def\references{\r@ferences\def\endmode{\r@ferr\par\endgroup}}

\let\endr@ferences=\endreferences
\def\endreferences{\r@fcurr=0
  {\loop\ifnum\r@fcurr<\r@fcount
    \advance\r@fcurr by 1\relax\expandafter\r@fis\expandafter{\number\r@fcurr}%
    \csname r@ftext\number\r@fcurr\endcsname%
  \repeat}\gdef\r@ferr{}\endr@ferences}


\let\r@fend=\endpaper\gdef\endpaper{\ifr@ffile
\immediate\write16{Cross References written on []\jobname.REF.}\fi\r@fend}

\catcode`@=12

\citeall\refto		
\citeall\ref		%
\citeall\Ref		%

\catcode`@=11
\newcount\tagnumber\tagnumber=0

\immediate\newwrite\eqnfile
\newif\if@qnfile\@qnfilefalse
\def\write@qn#1{}
\def\writenew@qn#1{}
\def\w@rnwrite#1{\write@qn{#1}\message{#1}}
\def\@rrwrite#1{\write@qn{#1}\errmessage{#1}}

\def\taghead#1{\gdef\t@ghead{#1}\global\tagnumber=0}
\def\t@ghead{}

\expandafter\def\csname @qnnum-3\endcsname
  {{\t@ghead\advance\tagnumber by -3\relax\number\tagnumber}}
\expandafter\def\csname @qnnum-2\endcsname
  {{\t@ghead\advance\tagnumber by -2\relax\number\tagnumber}}
\expandafter\def\csname @qnnum-1\endcsname
  {{\t@ghead\advance\tagnumber by -1\relax\number\tagnumber}}
\expandafter\def\csname @qnnum0\endcsname
  {\t@ghead\number\tagnumber}
\expandafter\def\csname @qnnum+1\endcsname
  {{\t@ghead\advance\tagnumber by 1\relax\number\tagnumber}}
\expandafter\def\csname @qnnum+2\endcsname
  {{\t@ghead\advance\tagnumber by 2\relax\number\tagnumber}}
\expandafter\def\csname @qnnum+3\endcsname
  {{\t@ghead\advance\tagnumber by 3\relax\number\tagnumber}}

\def\equationfile{%
  \@qnfiletrue\immediate\openout\eqnfile=\jobname.eqn%
  \def\write@qn##1{\if@qnfile\immediate\write\eqnfile{##1}\fi}
  \def\writenew@qn##1{\if@qnfile\immediate\write\eqnfile
    {\noexpand\tag{##1} = (\t@ghead\number\tagnumber)}\fi}
}

\def\callall#1{\xdef#1##1{#1{\noexpand\call{##1}}}}
\def\call#1{\each@rg\callr@nge{#1}}

\def\each@rg#1#2{{\let\thecsname=#1\expandafter\first@rg#2,\end,}}
\def\first@rg#1,{\thecsname{#1}\apply@rg}
\def\apply@rg#1,{\ifx\end#1\let\next=\relax%
\else,\thecsname{#1}\let\next=\apply@rg\fi\next}

\def\callr@nge#1{\calldor@nge#1-\end-}
\def\callr@ngeat#1\end-{#1}
\def\calldor@nge#1-#2-{\ifx\end#2\@qneatspace#1 %
  \else\calll@@p{#1}{#2}\callr@ngeat\fi}
\def\calll@@p#1#2{\ifnum#1>#2{\@rrwrite{Equation range #1-#2\space is bad.}
\errhelp{If you call a series of equations by the notation M-N, then M and
N must be integers, and N must be greater than or equal to M.}}\else%
 {\count0=#1\count1=#2\advance\count1
by1\relax\expandafter\@qncall\the\count0,%
  \loop\advance\count0 by1\relax%
    \ifnum\count0<\count1,\expandafter\@qncall\the\count0,%
  \repeat}\fi}

\def\@qneatspace#1#2 {\@qncall#1#2,}
\def\@qncall#1,{\ifunc@lled{#1}{\def\next{#1}\ifx\next\empty\else
  \w@rnwrite{Equation number \noexpand\(>>#1<<) has not been defined yet.}
  >>#1<<\fi}\else\csname @qnnum#1\endcsname\fi}

\let\eqnono=\eqno
\def\eqno(#1){\tag#1}
\def\tag#1$${\eqnono(\displayt@g#1 )$$}

\def\aligntag#1$${\gdef\tag##1\\{&(\displayt@g##1 )\cr}\eqalignno{#1\\}$$
  \gdef\tag##1$${\eqnono(\displayt@g##1 )$$}}

\def\eqalignno#1{\displ@y \tabskip\centering
  \halign to\displaywidth{\hfil$\displaystyle{##}$\tabskip\z@skip
    &$\displaystyle{{}##}$\hfil\tabskip\centering
    &\llap{$\displayt@gpar##$}\tabskip\z@skip\crcr
    #1\crcr}}

\def\displayt@gpar(#1){(\displayt@g#1 )}

\def\displayt@g#1 {\rm\ifunc@lled{#1}\global\advance\tagnumber by1
        {\def\next{#1}\ifx\next\empty\else\expandafter
        \xdef\csname @qnnum#1\endcsname{\t@ghead\number\tagnumber}\fi}%
  \writenew@qn{#1}\t@ghead\number\tagnumber\else
        {\edef\next{\t@ghead\number\tagnumber}%
        \expandafter\ifx\csname @qnnum#1\endcsname\next\else
        \w@rnwrite{Equation \noexpand\tag{#1} is a duplicate number.}\fi}%
  \csname @qnnum#1\endcsname\fi}

\def\ifunc@lled#1{\expandafter\ifx\csname @qnnum#1\endcsname\relax}

\let\@qnend=\end\gdef\end{\if@qnfile
\immediate\write16{Equation numbers written on []\jobname.EQN.}\fi\@qnend}

\catcode`@=12

\preprintno{IN92003, UCSBTH--92--37}
\title Critical Exponents of the Chiral Potts Model
from Conformal Field Theory
\author John L. Cardy$^*$
\affil Isaac Newton Institute for Mathematical Sciences
20 Clarkson Road
Cambridge CB3 0EH, UK
\abstract
The $Z_N$-invariant
chiral Potts model is considered as a perturbation of a
$Z_N$ conformal field theory.
In the self-dual case the renormalization group equations become
simple, and yield critical exponents and anisotropic scaling which agree
with exact results for the super-integrable lattice models. The continuum
theory is shown to possess an infinite number of conserved charges on
the self-dual line, which remain conserved when the theory is perturbed by
the energy operator.

\vfill
\noindent$^*$Permanent address: Department of Physics, University of
California, Santa Barbara CA 93106, U.S.A.
\eject

The $Z_N$ chiral Potts models, first introduced as models of systems
exhibiting \break
commensurate-incommensurate and melting transitions\refto{OST,HOW},
have more recently
become of interest as exactly solvable lattice models whose Boltzmann
weights satisfy the star-triangle relation, yet do not enjoy the so-called
\lq difference property' possessed by previously studied such models\refto{
GEN}.
The chiral Potts models are also unusual in that they are intrinsically
anisotropic in their critical properties\refto{CRIT,BAX2}, unlike most other
such
lattice
critical systems in which rotational symmetry is recovered in the scaling
limit, so that they are described by a \cft.

A special case of the chiral Potts model, the so-called superintegrable
case, has proved to be especially tractable\refto{HOW,BAX2,GR,MCOY}.
It becomes critical at the self-dual point, and the critical exponents
describing the specific heat singularity and the anisotropic scaling of the
correlation lengths have been obtained exactly: $\a=1-2/N$, $\nu_1=1$,
and $\nu_2=2/N$\refto{CRIT,BAX2}.
In addition, the presumably exact
conjectures $\b_j=j(N-j)/2N^2$
have been made\refto{HOW,CRIT,HENKEL} for the exponents
governing the order
parameters $\ev{e^{ij\th}}\sim (-t)^{\b_j}$.
Interestingly enough, the values found for the \lq thermodynamic' exponents
$\a$ and $\b_j$ agree with those of the
isotropic $Z_N$ models\refto{FZ1}
obtained when the chirality parameter is set to zero,
whose continuum limit at criticality is believed to correspond to the $Z_N$
\cfts\ first
studied by Fateev and Zamolodchikov\refto{FZ2}.
{}From the \rg\ point of view, this result is puzzling, since,
as will be discussed below, the chiral perturbation of the isotropic model
appears to be relevant, and therefore should be expected to
modify the critical exponents in an essential way.

In this paper, we give a simple explanation of this apparent paradox,
showing why the thermodynamic exponents are not changed.
In addition, we give a simple derivation of the anisotropic scaling
exponents $\nu_1$ and $\nu_2$. Our methods are
strictly valid only for weak chirality, close to the isotropic point. However,
they suggest that the exponents are in fact independent of the magnitude
of the chirality parameter, to all orders in perturbation theory.
The main non-universal feature, which appears to vary continuously along the
critical line, is the orientation of the principal axes with respect to
which anisotropic scaling occurs.
In addition, we show that there is a two-dimensional parameter space
in which these models are integrable in the continuum sense,
that is, they possess an infinite number of conserved charges.

In the simplest case, $N=3$, the reduced lattice Hamiltonian for this model
on a square lattice with sites labelled by pairs of integers $(x,y)$
is
$$
S=-\sum_{x,y}\left(K_x\cos(\D_x\th_{x,y}-\delta_x)
+K_y\cos(\D_y\th_{x,y}-\delta_y)\right)
\qquad,\eqno(S)
$$
where the $\th_{x,y}$ are angles which are integer multiples of $2\pi/3$,
and $\D_\mu$ ($\mu=x,y$) is the lattice difference operator.
The original version of this model, due to Ostlund\refto{OST}, corresponds
to $\delta_y=0$.
The self-dual case, of interest here, occurs for $\delta_y=\pm i\delta_x$.
The superintegrable point then corresponds to $\delta_x=\frac\pi6$.
When $\d_x=\d_y=0$, this model is the ordinary 3-state Potts model, whose
continuum limit at the critical point is described by the $Z_3$-invariant
\cft\ with central charge $c=\frac45$. When the chiral perturbation is turned
on, to which operators in the continuum theory does it couple? On the lattice,
$\d_\mu$ couples to $\sin\D_\mu\th$.
This transforms under $90^{\circ}$ lattice rotations like
a vector, and so should, in the continuum limit where full rotational
invariance
is recovered, transform as a combination of operators with spins
$\pm(4n+1)$, where $n=0,1,\ldots$.
The most relevant such operators of the $Z_3$ \cft, which are not total
derivatives, have spin $\pm1$, and (labelled by their scaling dimensions
$(h,\hb)$) are
\def\CH{\Phi_{(\frac75,\frac25)}}
\def\CHB{\Phi_{(\frac25,\frac75)}}
$\CH$ and $\CHB$. Note that they have
overall scaling dimension $x=\ffrac95<2$. This means that they are relevant
in the \rg\ sense. Considerations of rotational symmetry then imply that
$\d=\d_x+i\d_y$ couples to $\CH$ and $\db=\d_x-i\d_y$ to $\CHB$.

We now consider the duality properties of these operators.
The duality transformation involves three steps: first the Boltzmann weights
on each link are Fourier transformed with respect to $\D_\mu\th$, introducing
new integer-valued conjugate link variables $n_\mu$, then the sums over the
original variables $\th_{x,y}$ are carried out, which give constraints
of the form $\D_\mu n_\mu=0\,({\rm mod}\ 3)$ at each lattice site.
These are then solved
by writing $n_\mu\propto\e_{\mu\nu}\tilde\th_\nu$, where the
$\tilde\th_\nu$ are angles defined on the dual lattice, and $\e_{\mu\nu}$
is the totally antisymmetric symbol. Since, in the
Fourier transform, $n_\mu$ couples linearly to $i\d_\mu$, it follows that
$\tilde\th_\nu$ will couple to $i\e_{\nu\mu}\d_\mu$. Keeping track of the
factors more carefully, we find that under duality,
$$
\d_\mu\to i\e_{\mu\nu}\d_\nu   \qquad,\eqno(DUAL)
$$
so that $\d\to\d$ and $\db\to-\db$.
The model with $\db=0$, $\d\not=0$ is thus self-dual.
Equivalently we may say that, under duality, $\CH\to\CH$,
while\footnote*{This apparent
lack of symmetry between $\CH$ and $\CHB$ can be traced to a particular
choice for $\e_{\mu\nu}$. In fact, there are two possible duality
transformations
corresponding to each choice of sign. Under the other kind, it is $\CHB$
which is even.} $\CHB\to-\CHB$.
This is consistent with the fusion rules of these operators,\break
$\CH\cdot\CHB\sim\Phi_{(\frac25,\frac25)}+\cdots$, since the leading thermal
operator
$\Phi_{(\frac25,\frac25)}$ is odd under duality.

Thus we are led to consider the field theory described by the action
$$
S=S_0+\d\int\CH d^2\!x+\db\int\CHB d^2\!x+\t\int\Phi_{(\frac25,\frac25)}d^2\!x
\qquad,\eqno(ACTION)
$$
where $S_0$ is the action of the $Z_3$ \cft.
The self-dual subspace, with which we shall
be mostly concerned, then corresponds to $\db=\t=0$. We are therefore
considering a \cft\ perturbed by an irreducible spin one operator. This will
turn out to have greatly simplifying consequences.

Let us consider the \rg\ equation for the dimensionless coupling
$g=\d a^{1/5}$, reflecting the flow of this coupling under changes in the
microscopic lattice cut-off $a$.
It is conventional to work within a scheme where the lattice cut-off is
replaced by one which respects the rotational symmetry of the theory.
In particular, we may consider an expansion of the partition function in terms
of the connected
correlation functions of $\CH$, with an
ultraviolet regulator which prohibits
any two of the integration arguments approaching more closely than $a$.
Within such a scheme, the coefficients of the \rg\ $\b$-function to any finite
order in perturbation theory are given by integrals over correlation functions
evaluated in the unperturbed theory, which has rotational symmetry.
This has the consequence that, to all orders
in perturbation theory, the \rg\ equation
for $g$ is given by its first term only:
$$
\dot g=\ffrac15g  \qquad,\eqno(GDOT)
$$
because any higher order terms $g^n$ with $n>1$ would not transform correctly
under rotations. Explicit calculations at low orders confirm this result.
In the same way, if we consider any scalar perturbation of the action $S$,
for example the thermal operator, then its \rg\ equation can contain no
dependence
on $g$, so that
$$
\dot\t=\ffrac65\t +O(\t^3)  \qquad.\eqno(TDOT)
$$
The same is true for all
other such scalar quantities which could enter the free energy.

The \rg\ equations, within
such a scheme, are therefore trivial.
There are only two possible fixed points, $g=0$ (the ordinary
3-state Potts model) and $g\to\infty$, to which all theories with
$g\not=0$ flow. Note that
this fixed point lies at infinity in a rotationally invariant regularization
scheme, but in others, for example a lattice scheme where terms $O(g^{4n+1})$
may appear on the \rhs\ of \(GDOT), it may move into a finite value.
The second equation \(TDOT) implies that the \rg\ eigenvalues of all scalar
operators are the same at $g=0$ and $g\to\infty$.
This has the consequence that the exponents governing the
singular behaviour of the free energy as
a function of temperature, magnetic field, and so on
are the same. It is consistent with the observation that they agree
also at the superintegrable point.

Next, we discuss the anisotropic scaling of correlation functions.
This is a consequence of the appearance of
a novel kind of space-time symmetry, which is most easily
seen through the properties of the
stress tensor. The argument is a simple generalization of that
given by Zamolodchikov\refto{ZAM} for perturbations of a \cft\ by a scalar
operator.
At the conformal point $\d=0$ there are two
non-zero components $T=T_{zz}$ and $\Tb=T_{\zb\zb}$, where $\partial_{\zb}T=
\partial_z\Tb=0$.
The other components
$T_{z\zb}$ and $T_{\zb z}$ vanish. When $\d\not=0$,
this is no longer true. A correlation function $\ev{T(z)\ldots}$
is modified to first order in $\d$ by a term
$$
-\d\int\ev{T(z)\CH(z_1,\zb_1)\ldots\ldots}d^2\!z_1
\qquad.\eqno()
$$
Using the \ope
$$
T(z)\cdot\CH(z_1,\zb_1)={\ffrac75\over(z-z_1)^2}\,\CH(z,\zb)+{1-\ffrac75\over
z-z_1}
\,\partial_z\CH(z,\zb)+\cdots
\quad,\eqno()
$$
and the fact that $\partial_{\zb}(z-z_1)^{-1}=\ffrac\pi2\d^{(2)}(z-z_1)$,
it follows that
$$
\partial_{\zb}T=\ffrac\pi2(1-\ffrac75)\,\d\,\CH + O(\d^2)
\qquad.\eqno()
$$
Now we should expect to be able to express all the terms on the \rhs\ in terms
of the renormalized coupling constant and renormalized operators of the \cft.
But \(GDOT) implies that $\d$ needs no renormalization (as may be checked,
there
are no new ultraviolet divergences in correlation functions), and a simple
dimensional argument shows that the operator multiplying the term of order
$\d^n$ would have scaling dimensions $(2+2n/5,1-3n/5)$. Since these must
all be non-negative, only the
$n=1$ term is in fact present.
It then follows from the conservation equation $\partial_{\zb}T+\partial_z
T_{\zb z}
=0$ that
$$
T_{\zb z}=-\ffrac\pi2(1-\ffrac75)\,\d\,\CH \qquad,\eqno(ZBZ)
$$
and similarly that
$$
T_{z\zb}=-\ffrac\pi2(1-\ffrac25)\,\d\,\CH  \qquad.\eqno(ZZB)
$$
The fact that these are not equal implies that rotational symmetry is broken,
as expected. Instead we see that
$$
T_{\zb z}=-\frac23T_{z\zb}   \qquad.\eqno(23)
$$
This implies the existence of an unusual symmetry, under transformations
which are mixtures of rotations and dilatations
$$
z\to e^{3\om}z\quad,\qquad\zb\to e^{2\om}\zb  \quad,\eqno(NEWSYMM)
$$
where $\om$ is arbitrary.
However, we see immediately that these transformations do not preserve
the reality of $z+\zb$ and $i(z-\zb)$, and therefore do not appear to
be physical. This requires careful discussion.
Since $\d_y$ is pure imaginary, the Boltzmann weights are not necessarily real
and positive. This means that there is no straightforward interpretation
as a statistical model. Several of the usual properties of such systems,
such as insensitivity of bulk thermodynamic quantities to boundary conditions,
can fail in this case, and it appears that the usual arguments supporting the
existence of a continuum limit may not go through either. However, as pointed
out by several authors, although the transfer matrix corresponding to \(S) is
not real, it is hermitian, as is the quantum hamiltonian which arises in the
limiting case. Therefore such a transfer matrix, or quantum hamiltonian, should
define a sensible quantum theory in $1+1$ dimensions with the possibility of
a continuum limit as a quantum field theory.\footnote*{These observations
do not invalidate our previous assertions about the structure of the \rg\
equations,
which were made within the context of perturbation theory, which is
well-defined
in Euclidean space. However, it should also be possible to work entirely in
real
time, and require a perturbative regularization procedure which respects
Lorentz invariance.}
Within this picture the symmetry
\(NEWSYMM) is now physical,
since in real time $z$ and $\zb$ get replaced by light-cone
co-ordinates $x^{\pm}=t\pm x$,
while \(23) becomes $T_{-+}=\frac23T_{+-}$.
The generator of this symmetry is $\ffrac52D+\ffrac12M$, where $D$ is the
generator of
dilatations, and $M$ that of Lorentz boosts.

This has important consequences for the spectrum.
If the hamiltonian and momentum operators are denoted by $H$ and $P$,
and we introduce their light-cone combinations $P^{\pm}=H\pm P$, then it
follows from the commutation relations of $H$, $P$, $D$ and $M$ that
if $\ket{p^+,p^-}$ is an eigenstate, so is $\ket{e^{-2\om}p^+,e^{-3\om}p^-}$.
\def\vac{\Psi_0}
The density of states is constant along curves $p^-=\k\d^{-5/2}(p^+)^{3/2}$,
where $\k$ is constant.
Within perturbation theory in $\d$, only states with positive $p^+$ and
$p^-$, and therefore positive $\k$, may arise, since the unperturbed
correlation
functions have support only inside the light-cone.
However, this does not rule out the possibility of non-perturbative states
with $\k<0$. For the spectrum of $H$
to be bounded from below, there must be a minimum value of $\k$, say
$\k_{\rm min}$. If $\k_{\rm min}<0$,
the ground state will then lie on this curve, and will have non-zero eigenvalue
of $P$, implying a spontaneous breaking of translation symmetry.
In fact, exact calculations at the superintegrable point\refto{MCOYXING},
and numerical
calculations for more general values of the chiral parameter\refto{VG},
indicate that
this in fact happens. As the critical line is approached, a whole sequence of
first-order transitions takes place in which states with non-zero momentum
cross the lowest energy state in the zero-momentum sector.
The results for the critical exponents, however, refer to quantities
evaluated in the perturbative sector, whose lowest energy state
we denote by $\ket\vac$.

A two-point correlation function of an arbitrary scalar operator $\phi$
can be written as an integral over a spectral density
$$
\bra\vac T[\phi(x^+,x^-)\phi(0,0)]\ket\vac
=\int\rho(p^+,p^-)e^{ip\cdot x}dp^+dp^-
\qquad,\eqno(RHO)
$$
where
$$
\rho(p^+,p^-)=\sum_n|\ev{\Psi_0|\phi|n}|^2\d(p^+_n-p^+)\d(p^-_n-p^-)
\eqno()
$$
which will have the scaling form
$$
\rho(p^+,p^-)=(p^+p^-)^{x_\phi-1}\,f\left({\d(p^+)^{2/5}\over(p^-)^{3/5}}\right)
\qquad,\eqno(SCA)
$$
where we have used the facts that $\phi$ transforms under dilatations according
to its scaling dimension $x_\phi$, and is a Lorentz scalar, and that $\ket\vac$
is annihilated by $P^+$ and $P^-$.

All this is supposed to be valid on the self-dual line. When the thermal
perturbation is added, the scaling function $\Phi$ will also depend on the
additional scaling variable $\t(p^+p^-)^{-3/5}$ (where the exponent $\frac35$
reflects the value of the
thermal eigenvalue.) We may write this two-variable
scaling form alternatively as
$$
\rho(p^+,p^-)=(p^+p^-)^{x_\phi-1}\,\tilde f\left(p^+\xi^-,p^-\xi^+\right)
\qquad,\eqno(RHPSCA)
$$
where
$$
\eqalign{
\xi^+&\sim\d\t^{-1}\quad,\cr
\xi^-&\sim\d^{-1}\t^{-2/3}\quad.\cr}  \eqno(NU)
$$
We therefore identify the anisotropic correlation length exponents
$\nu_1=1$ and $\nu_2=2/3$.

It will be noticed immediately that the principal axes with respect to
which this anisotropic scaling takes place are $x^{\pm}$, not $t$ and $x$
as is found in studies of the superintegrable case\refto{HOW,CRIT}. We now
offer an
explanation of this. The lattice theory differs from the continuum action
\(ACTION) by terms which break rotational, or Lorentz, symmetry.
These terms are irrelevant in the isotropic theory at the critical point,
since they correspond to operators of spin 4 or higher. When such an
isotropic critical theory is perturbed by a scalar operator, it is generally
true that such operators remain irrelevant, at least in the scaling
region. This is consistent with the observation that no new terms in their
renormalization arise from the scalar perturbation. When a theory is
perturbed by a non-scalar operator, however, this is no longer valid.
Indeed, in the present case the coupling constant $g_s$ of an operator of
spin $s$ can be renormalized at order $g^s$ by the interaction, so that its
\rg\ equation has the form
$$
\dot g_s=y_sg_s+\const g^s+\cdots
\qquad.\eqno()
$$
A particular case of interest is the $(zz)$ component of the stress tensor
$T$ whose coupling $g_2$ will get a contribution at order $g^2$. In fact,
this may be computed
from the \ope\ $\CH\cdot\CH\sim\frac72T+\cdots$ to give
$$
\dot g_2= \ffrac{7\pi}2 g^2+\cdots
\qquad.\eqno(G2DOT)
$$
The stress tensor is an example of a redundant operator. When added to the
action, it does not modify the critical exponents, but instead
its effect may be removed by an appropriate co-ordinate transformation.
In our case, this is accomplished by sending\footnote*{The \ope\ coefficient
$\frac72$ comes from the scaling dimension $\frac75$ normalized by $\frac c2=
\frac25$. The extra factor of $2\pi$ in \eq{20} arises from that in the
conventional definition of the stress tensor.}
$$
x^+\to x^++7\pi^2g^2\,x^-
\qquad.\eqno()
$$
There are, in addition, higher order effects in $a$ which come from
terms in \(G2DOT) like $g_{-4}g^6$, and so on.
All of these will be non-universal, and cannot
be calculated within our continuum approach.
Note that all these additional operators do not affect the simple \rg\ equation
\(GDOT) for $g$, since they are all have even spin.
Although this shows that non-universal lattice effects may rotate
the principal scaling axes away from the light-like directions $x^\pm$,
it does not explain why the axes should be precisely parallel to
$t$ and $x$ at the
superintegrable point. Perhaps there is a hidden symmetry at this point which
requires such an alignment.

Although the arguments above have been presented for $N=3$ for clarity, they
generalize straightforwardly to arbitrary $N$. In that case, in the
isotropic $Z_N$ model, the leading thermal operator has scaling dimensions
$(2/(N+2),2/(N+2))$, and the leading spin 1 operator
$((N+4)/(N+2),2/(N+2))$. We take this as coupling
to the chiral parameter $\d$. All the previous arguments then go through, and
in
particular we find that $T_{-+}=\ffrac2NT_{+-}$. This leads to anisotropic
correlation length exponents $\nu_1=1$, $\nu_2=2/N$, in agreement with
the exact results found at the superintegrable point\refto{CRIT,BAX2}.

Finally, we give a simple argument that these continuum models should
be integrable.
In the conformal field theory, there is an infinite number of conserved charges
corresponding to holomorphic and antiholomorphic currents $(T^{(s)}(z),0)$
and $(0,\Tb^{(s)}(\zb))$ of arbitrarily high spin $s$. When the theory is
perturbed by the thermal
operator $\Phi_{2/N+2,2/N+2}$, it is well known that an infinite subset of
these remain conserved, with spins $s$ following the exponents of
of $A_{N-1}$ modulo the Coxeter number\refto{CONS}. This may be traced to the
property that the residue at $z_1=z$ in the \ope\ of $T^{(s)}(z)$ and
$\Phi_{2/N+2,2/N+2}(z_1,\zb_1)$ is a total derivative with respect to
$z$\refto{
ZAM}.
Similarly, an infinite set of antiholomorphic currents remain conserved for
an analogous reason.
Now all of the $Z_N$ \cfts\ possess an extended $W$-symmetry, generated
by conserved currents $W^{(n)}(z)$ and $\Wb^{(n)}(\zb)$\refto{WSYMM}.
With respect to
this extended symmetry, the chiral fields of dimension $((N+4)/(N+2),
2/(N+2))$ are not primary, but $W$-descendents of the form
$W_{-1}^{(n)}\Phi_{2/N+2,2/N+2}$ for some $n$. Since the holomorphic
current $W^{(n)}$ commutes with all the antiholomorphic ones, it follows
that all the antiholomorphic currents which remain conserved under a
purely thermal perturbation remain conserved when the chiral perturbation is
also present. In general, however, the holomorphic currents are
no longer conserved.
It is unclear at this point what consequences may be drawn from this.
The theory with both types of perturbation is expected to be massive, and
therefore to possess a sensible $S$-matrix. However, in the absence of
Lorentz invariance (which holds only when $\d=0$), or the mixed symmetry
discussed earlier (which holds only when $\t=0$), there is no simple way of
determining how these conserved charges act on the asymptotic states.

The manifolds of integrability are the planes $\db=0$ and $\d=0$
in the space parametrized by $(\t,\d,\db)$.
These appear to differ from the lattice case, where, close
to the isotropic point, the exact solution manifold, expressed in our
coordinates, has the form
$\t\sim\d\db$. It is possible that irrelevant couplings are responsible
for this difference, or that continuum and lattice
integrability are rather different properties.

In passing, we comment on the more physical case of of the uniaxial model
with $\d_x\not=0$, $\d_y=0$. In this case symmetry arguments are
are far less powerful. The \rg\ equation for $\d_x$ is no longer trivial,
and gets corrections at order $\d_x^5$.
These are very difficult to compute, involving as they do an integral
over a 6-point function. However, they do suggest the possibility of a
non-trivial fixed point for $\d_x\not=0$. If this were to occur, it would
correspond to a Lifshitz point, whose existence is still
controversial in numerical studies of the $N=3$ model\refto{LIF}.

In summary, we have given simple symmetry arguments for the values
of the critical exponents of the $Z_N$ chiral Potts models on the self-dual
line. These arguments
are valid in the limit of weak chirality, but the results are independent
of the its magnitude, and are consistent with those found at the
superintegrable
point. We have argued that the continuum theory makes sense only in
Minkowski space, and found that it exhibits an unusual space-time symmetry,
which allows for the observed spontaneous breaking of translational
symmetry in the true ground state.
Our analysis indicates that, while the anisotropic correlation length
exponents are universal, the principal axes to which they refer are not, and
that these should rotate away from the light-like directions as the chirality
parameter $\d$ is increased.
This may explain the inconclusive results obtained for mass-gap scaling in
numerical
work for small values of $\d$\refto{HOW}.
Crossover theory suggests that the asymptotic
regime should be reached only for reduced temperatures of the order of
$\d^{2N/(N-2)}$, which will  be a very narrow region, especially for $N=3$.
However, even then, the exponent
$\nu_1=1$ should be observed only if the gap to states with the correct
non-zero
value of the momentum is considered. In all other cases, the other exponent
$\nu_2=2/N$ should be observed. It would be interesting to see whether
the predicted rotation of the scaling axes occurs in the
vicinity of the superintegrable point, where numerical work should be more
conclusive.

The author thanks R.~Baxter and B.~McCoy for useful discussions on the
lattice results, and the Isaac Newton Institute for Mathematical
Sciences for its hospitality. This work was supported by the Isaac Newton
Institute, the UK Science and Engineering Research Council, and by
the US National Science Foundation Grant PHY 91-16964.

\references

\refis{OST} S. Ostlund, \journal Phys. Rev. B, 24, 398, 1981.

\refis{HOW} S.~Howes, L.~P.~Kadanoff and M.~den~Nijs, \np B 257[FS7], 169,
1983.

\refis{GEN} H.~Au-Yang, B.~M.~McCoy, J.~H.~H.~Perk, S.~Tang and M.-L.~Yan,
\journal Phys. Lett. A, 123, 219, 1987; B.~M.~McCoy, J.~H.~H.~Perk, S.~Tang,
\journal Phys. Lett. A, 125, 9, 1987; R.~J.~Baxter, J.~H.~H.~Perk and
H.~Au-Yang, \journal Phys. Lett. A, 128, 138, 1988.

\refis{CRIT} G.~Albertini, B.~M.~McCoy, J.~H.~H.~Perk and S.~Tang,
\np B 314, 741, 1989.

\refis{GR} G.~v.~Gehlen and V.~Rittenberg, \np B 257[FS14], 351, 1985.

\refis{MCOY} G.~Albertini, B.~M.~McCoy and J.~H.~H.~Perk, \journal
Phys. Lett. A, 135, 159, 1989; \journal Phys. Lett. A, 139, 204, 1989;
\journal Advanced Studies in Pure Math., 19, 1, 1989.

\refis{BAX2}  R.~J.~Baxter, \journal Phys. Lett. A, 133, 185, 1988;
 R.~J.~Baxter, \journal J. Stat. Phys., 57, 1, 1989.

\refis{HENKEL} M.~Henkel and J.~Lacki, \journal Phys. Lett. A, 138, 105, 1989.

\refis{FZ1} V.~A.~Fateev and A.~B.~Zamolodchikov, \journal Phys. Lett. A, 92,
37, 1982.

\refis{FZ2} V.~A.~Fateev and A.~B.~Zamolodchikov, \journal Sov. Phys. JETP,
62, 215, 1985.

\refis{MCOYXING} G.~Albertini, B.~M.~McCoy and J.~H.~H.~Perk, \journal
Advanced Studies in Pure Math., 19, 1, 1989; G.~Albertini and B.~M.~McCoy,
\np B 350, 745, 1991.

\refis{VG} G.~v.~Gehlen, proceedings of International Symposium on Advanced
Topics of Quantum Physics, Taiyuan, China, 1992; Bonn preprint HE--92--18.

\refis{ZAM} A.~B.~Zamolodchikov, \journal Int. J. Mod. Phys., A3, 743, 1988.

\refis{CONS} V.~A.~Fateev, \journal Int. J. Mod. Phys., A6, 2109, 1991.

\refis{WSYMM} V.~A.~Fateev and S.~I.~Lukyanov, \journal Int. J. Mod. Phys.,
A3, 507, 1988.

\refis{LIF} P.~Centen, V.~Rittenberg  and M.~Marcu, \np B 205, 585, 1982;
G.~v.~Gehlen and V.~Rittenberg, \np B 230, 455, 1984;
T.~Vescan, V.~Rittenberg and G.~v.~Gehlen, \journal J. Phys. A, 19, 1957, 1986;
M.~Siegert and H.~U.~Everts, \journal J. Phys. A, 22, 783, 1989;
H.~U.~Everts and H.~R\"oder, \journal J. Phys. A, 22, 2475, 1989.

\endreferences

\endit